\documentclass[12pt,a4paper]{article}
\usepackage{amsmath}
\usepackage{graphics}

\newcommand{\hepth}[1]{{\tt hep-th/#1}}
\newcommand{\plb}[3]{Phys.Lett. {\bf B#1} (#2) #3}
\newcommand{\npb}[3]{Nucl.Phys. {\bf B#1} (#2) #3}

\newcommand{\ep}{\varepsilon} 
\newcommand{\lam}{\lambda}
\renewcommand{\th}{\Theta}
\newcommand{\bl}[1]{\left(#1\right)}
\newcommand{\nn}{\nonumber}

\newcommand{\im}[1]{\text{Im}(#1)}
\advance\textheight by \topskip
\makeatletter

\@addtoreset{equation}{section}
\def\section{\@startsection {section}{1}{\z@}{-8.5ex plus -1ex minus
 -.2ex}{3.3ex plus .2ex}{\large\bf\centering}}
\def\subsection{\@startsection{subsection}{2}{\z@}{-3.25ex plus
 -1ex minus -.2ex}{1.5ex plus .2ex}{\bf}}
\def\subsubsection{\@startsection{subsubsection}{3}{\z@}{-3.25ex plus%
 -1ex minus -.2ex}{1.5ex plus .2ex}{\sl}}

\begin{document}
\begin{titlepage}
\vspace*{-2cm}
\begin{flushright}

DTP--00/67\\ hep-th/0008237 \\

\end{flushright}

\vspace{0.3cm}

\begin{center}
{\Large {\bf Reflection factors and a two-parameter family of boundary
bound states in the sinh-Gordon model  }}\\ \vspace{1cm} {\large \bf E.\
Corrigan\footnote{\noindent E-mail: {\tt ec9@york.ac.uk}} and A.\
Taormina\footnote{\noindent E-mail: {\tt anne.taormina@durham.ac.uk}} }\\
\vspace{0.3cm} {${}^a$}\em Department of Mathematics \\ University of York\\
York YO10 5DD, U.K.\\ \vspace{0.3cm} {${}^b$\em \it Department of Mathematical
Sciences\\ University of Durham\\ Durham DH1 3LE, U.K.}\\ \vspace{1cm}
{\bf{ABSTRACT}}
\end{center}

\begin{quote}
 The investigation of boundary breather states of the sinh-Gordon model
restricted to a half-line is revisited. Properties of the classical boundary
breathers for the two-parameter family of integrable boundary conditions are
reviewed and extended. The energy spectrum of the quantized boundary
states is computed,
firstly by using a bootstrap technique and, subsequently using a WKB
approximation. Requiring that the two descriptions of the spectrum agree with
one another allows a determination of the relationship between the boundary
parameters, the bulk coupling constant, and the two parameters appearing in
the reflection factor describing the scattering of the sinh-Gordon particle
from the boundary. These calculations had been performed previously
for the case in which the boundary conditions preserve the
bulk $Z_2$ symmetry of the model. The significantly more difficult case of
general boundary conditions which violate the bulk symmetry
is treated in this article. The results
clarify the weak-strong coupling duality of the sinh-Gordon model with
integrable boundary conditions.
\end{quote}

\vfill
\end{titlepage}

\section{Introduction}
\label{s:intro}

In the study of the sinh-Gordon model restricted to a half-line by
integrable boundary conditions, an important issue is to determine how
the reflection factor depends upon the two parameters introduced at the
boundary. Some progress was recently made in tackling this question
\cite{Cor99, Chen00}. Namely, the idea of
\cite{Cor99} was to compute the bound-state spectrum of the model in two
different ways, firstly by using a boundary bootstrap principle, and
secondly by quantizing the classical boundary breather states using a
WKB approach
\cite{DHN75,Raj82}. Each method provides an independent description of
the energy spectrum of the boundary states, and comparing the two
provides information relating the boundary parameters and the bulk
coupling to the reflection factor. Clearly, one expects the most
general situation where the two boundary parameters are independent to
be technically more involved. It is therefore natural - as was done in
\cite{Cor99} - to first implement the programme outlined above in the
special case where the bulk symmetry $\phi\rightarrow -\phi$ is
preserved at the boundary, by requiring the two parameters to be equal.
A further step was taken in \cite{Chen00}, where a perturbative
calculation to lowest order in the bulk coupling and to
first order in the difference of the two boundary parameters was used
to make an informed guess as to the general dependence of the reflection
factor on those boundary parameters. The results of the present
paper, which extends the analysis in \cite{Cor99} to the case
where the two boundary parameters are different, will underpin
that guess. Finally, a version of a weak-strong coupling duality
transformation appropriate to the model with boundary conditions was
also proposed in \cite{Chen00} and the conclusions drawn from our paper
give further support to that conjecture.

Following a brief introduction to the sinh-Gordon model with integrable
boundary conditions in section 1, we recall the classical breather
solutions in section 2 and describe the properties and facts we need for
the semi-classical quantization. In section 3, we review the boundary
state bootstrap and use it to derive a formula for the boundary state
energy spectrum. The adapted Dashen-Hasslacher-Neveu method is described
in section 4 and used to provide an alternative calculation of the
boundary state spectrum in terms of the boundary parameters. Our results
and some additional remarks are summarized in section 5.

\section{The sinh-Gordon model on the half-line}

\label{s:sg}

The sinh-Gordon model describes a single real scalar field $\phi$
in 1+1 dimensions with exponential self-interaction. The field
equation is
\begin{equation}\label{fe}
\partial_t^2\phi-\partial_x^2\phi+
\frac{\sqrt{8}m^2}{\beta}\sinh(\sqrt{2}\beta\phi)=0,
\end{equation}
where $m$ and $\beta$ are parameters and we have used
normalizations customary in affine Toda field theories of which
the sinh-Gordon model is the simplest \cite{Bra90}. The
dimensional mass parameter $m$ will be set to unity for convenience.

In contrast to the sine-Gordon model, with its soliton and breather
solutions, the sinh-Gordon model is at first sight
 relatively uninteresting. There is
a constant vacuum solution $\phi=0$ and,
in the quantum theory, the small oscillations around this vacuum
correspond to the sinh-Gordon particle. In the bulk, the spectrum of
the model consists of a single species of scalar particle
interacting
with itself. Nevertheless, however simple the model may appear,
 it is not easy to
analyze it directly \cite{Skly89}, and much of what is known about it
has been deduced from features of the lightest breather in the
sine-Gordon theory.

The sinh-Gordon model is integrable which implies in particular that
there are
infinitely many mutually commuting,
independent conserved charges $Q_{\pm s}$, where $s$ is any
odd integer, and the S-matrix describing the scattering of two sinh-Gordon
particles with  relative rapidity $\th$ is conjectured to be given
 by \cite{Fadd78,
Zam79},
\begin{equation}\label{s}
 S(\th)=-\frac{1}{(B)_\th (2-B)_\th}.
\end{equation}
In \eqref{s} we have used the convenient block notation \cite{Bra90}
\begin{equation}\label{blo}
\bl{x}_\th=\frac{\sinh\left(\frac{\th}{2}+\frac{i\pi x}{4}\right)}
{\sinh\left(\frac{\th}{2}-\frac{i\pi x}{4}\right)},
\end{equation}
and the coupling constant $B$ is related to the bare coupling
constant $\beta$
by $B=2\beta^2/(4\pi+\beta^2)$. For compactness, we will generally
omit the subscript $\th$ from the block notation since it is generally
clear from the context what is intended.

The sinh-Gordon model can be restricted to the left half-line
$-\infty\leq x\leq 0$ without losing integrability by imposing the
boundary condition
\begin{equation}\label{bc}
 \left. \partial_x\phi\right|_0=\frac{\sqrt{2}m}{\beta}
 \left(\ep_0e^{-\frac{\beta}{\sqrt{2}}\phi(0,t)}-
\ep_1e^{\frac{\beta}{\sqrt{2}}\phi(0,t)}\right) ,
\end{equation}
where $\ep_0$ and $\ep_1$ are two additional parameters \cite{Ghosh94a,
MacI95}. Again, $m$ sets the scale but is taken to be  unity for
convenience in what follows. This set of boundary conditions generally
breaks the reflection
symmetry $\phi\rightarrow -\phi$ of the  model although the symmetry is
preserved when $\ep_0=\ep_1\equiv\ep$.

Assuming factorization, the description of the sinh-Gordon particles
on  the half line
does not only require the two-particle scattering amplitude \eqref{s},
but  also
the amplitude for the reflection of a single particle from the boundary. This
reflection amplitude was deduced from the lowest breather reflection amplitude
in the sine-Gordon model by analytic continuation in the coupling constant
(i.e., the continuation $\lam\rightarrow -2/B$ in the notation of
\cite{Ghosh94a}). Using the breather reflection amplitudes calculated by
Ghoshal \cite{Ghosh94b}, the analytic continuation leads
to \footnote{In Ghoshal's notation $E=B\eta/\pi, \
F=iB\vartheta/\pi$.}
\begin{equation}\label{br}
K_q(\theta,\ep_0,\ep_1, \beta)=\frac{(1)(2-B/2)(1+B/2)}
{(1-E(\ep_0,\ep_1,\beta ))(1+E(\ep_0,\ep_1,\beta ))
(1-F(\ep_0,\ep_1,\beta ))(1+F(\ep_0,\ep_1,\beta ))},
\end{equation}
where we are again using the block notation from \eqref{blo} but, here,
$\theta$ represents the rapidity of a single particle. When the
bulk reflection symmetry is preserved one of the two parameters $E$ or $F$
vanishes. Without loss of generality, we shall take the vanishing parameter
in the symmetric case  to be $F$.
All reflection factors satisfy the crossing-unitarity relation
which, in the case of scalar reflection factors, reads,
\begin{equation}\label{cu}
K_q\left(\theta + \frac{i\pi}{2}\right) K_q\left(\theta-\frac{i\pi}{2}
\right) S(2\theta)=1.
\end{equation}
Actually, \eqref{br} is the simplest solution to the crossing-unitarity
relation using the S-matrix \eqref{s}, taking into account the independently
calculated classical limit of the reflection factors. In \cite{Cor95} the
classical reflection factor was found to be given by the formula
\begin{equation}\label{bclass}
 K_c(\theta,\ep_0,\ep_1, \beta)=-\,
\frac{(1)^2}{(1-a_0-a_1)(1+a_0+a_1) (1-a_0+a_1) (1+a_0-a_1)},
\end{equation}
in which it was convenient to use an alternative expression for
the boundary parameters, viz.,
\begin{equation}\label{ea}
\epsilon_0=\cos\pi a_0, \quad \epsilon_1=\cos\pi a_1 .
\end{equation}
The formula is well-defined provided we select the ranges $0\le a_i\le
1$ for $i=0,1$, to ensure a one to one correspondence between the
alternative parameters.
Clearly, \eqref{br} has the correct limit provided
\begin{equation}\label{EFlimits}
E\rightarrow a_0+a_1,\qquad F\rightarrow a_0-a_1.
\end{equation}
Recently, on the basis of a perturbative calculation, the relationship
between the various parameters was conjectured to be \cite{Chen00},
\begin{equation}\label{EFconjecture}
E=(a_0+a_1)(1-B/2),\qquad F=(a_0-a_1)(1-B/2).
\end{equation}
The principal purpose of this paper is to provide further evidence for
these formulae.

It was noted in \cite{Cor99}  that contrary to the situation on the whole
line, the sinh-Gordon equation restricted to a half-line by integrable
boundary conditions has non-singular, finite energy, breather solutions. These
solutions were described in some detail in \cite{Cor99}, particularly for the
special case in which $\epsilon_0=\epsilon_1$. Here, we shall concentrate on
the solutions containing two independent boundary parameters. The
existence of these special solutions and their associated states makes
it clear that the sinh-Gordon model has a rather more interesting
structure than one would be led to believe on the basis of
bulk calculations.

\section{Boundary Breathers}

\label{s:bb}

The boundary breathers may be conveniently described  following Hirota's
prescription \cite{Hiro80}. Generally, provided we set
\begin{equation}\label{st}
 \phi=-\frac{\sqrt{2}}{\beta}\ln\frac{\tau_+}{\tau_-}\, ,
\end{equation}
and choose the two $\tau$-functions as follows:
\begin{equation}\label{taupm}
\tau_{\pm}=1\pm (E_1+E_2+E_3) + (A_{12}E_1E_2+A_{13}E_1E_3 + A_{13}E_1E_3) \pm
A_{12}A_{13}A_{23}E_1E_2E_3,
\end{equation}
with
\begin{equation}\label{taupieces}
E_p=e^{a_p x+b_p t +c_p}, \quad a_p=2\cosh\rho_p,\quad b_p =2\sinh \rho_p,
\quad A_{pq}=\tanh^2\left(\frac{\rho_p-\rho_q}{2}\right),
\end{equation}
then we have just enough freedom to accommodate the general boundary
conditions \eqref{bc}. In effect, with general boundary conditions, the
solutions have the flavour of a `breather' superimposed on a
stationary `soliton',
borrowing the language of the sine-Gordon model. In detail, it is enough to
choose $E_1$ and $E_2$ to be complex conjugate partners  and periodic
in $t$, and
to take $E_3$ to be real and independent of $t$; thus,
 setting $\rho_1=-\rho_2=i\rho$
and $\rho_3=0$, one has:
\begin{equation}\label{choices}
E_1=e^{2x\cos\rho+2it\sin\rho+c}=E_2^*, \quad E_3=e^{2x+d};
\quad A_{12}=-\tan^2\rho,
\ \ A_{13}=A_{23}=-\tan^2(\rho/2).
\end{equation}
The period  of the breather is $\pi/\sin\rho$.
To match the boundary conditions, $c$ and $d$ need to be determined. It is
convenient (actually just a change of origin in $t$) to take $c$ to be real.
Then, the boundary conditions \eqref{bc} are satisfied provided
\begin{equation}\label{cd}
e^c=\frac{s}{\tan\rho},\ \ \ e^d=\frac{r}{\tan^2(\rho/2)}
\end{equation}
where
\begin{eqnarray}\label{cda}
& &\hspace{3cm} r=\frac{\sin\frac{\pi a_0}{2}-\sin\frac{\pi a_1}{2} }{
\sin\frac{\pi a_0}{2}+\sin\frac{\pi a_1}{2}}\nonumber\\ & & \\
& &s^2=\frac{1+\cos\rho}{1-\cos\rho}\ \frac{\cos\frac{\pi(a_0+a_1)}{2}
+\cos\rho}{\cos\frac{\pi(a_0+a_1)}{2}
-\cos\rho}\ \frac{\cos\frac{\pi(a_0-a_1)}{2}
-\cos\rho}{\cos\frac{\pi(a_0-a_1)}{2}
+\cos\rho}.\nonumber
\end{eqnarray}
In \eqref{cda} we have made use of the alternative parametrisation
\eqref{ea}.
There are other ways of writing \eqref{cda} which are useful when it
comes to evaluating certain integrals. For example, setting
$q=\tan(\rho/2)$ and $q_{\pm}=\tan(\pi a_\pm/2)$, where
$a_\pm = (a_0\pm a_1)/2$, we have
\begin{equation}\label{cdaliter}
s^2 = \frac{1}{q^2}\ \frac{1-q^2q_+^2}{1-q^{-2}q_+^2}
\ \frac{1-q^{-2}q_-^2}{1-q^2q_-^2},\qquad
r=\frac{q_-}{q_+}.
\end{equation}

For symmetrical boundary conditions with $\ep_0=\ep_1=\ep$,
the terms containing $E_3$ are
 not required (effectively $d\rightarrow -\infty$), and the
other pieces of  the $\tau$-functions collapse to:
\begin{equation}\label{bt}
\tau_\pm=1\pm 2\cos(2t\sin\rho )e^{2x\cos\rho
}\frac{1}{\tan\rho} \sqrt{\frac{\ep+\cos\rho}{\ep-\cos\rho}}-
e^{4x\cos\rho }\left(\frac{\ep+\cos\rho}{\ep-\cos\rho}\right).
\end{equation}
The expression \eqref{bt} is quite easy to analyze.
In particular, the solutions
ought to be real and have no singularities in the region $x<0$.
It was pointed out in \cite{Cor99} that these requirements are met
provided  the parameters $\ep$ and $\rho$ are suitably
restricted:
\begin{equation}\label{co}
 -1<\ep<0~~~\text{ and }~~~\cos\rho<-\ep.
\end{equation}
It is interesting that the upper limit $\cos\rho =-\ep$ corresponds
to a choice of frequency at
which the amplitude of the solution has collapsed to zero yielding the
vacuum configuration, $\phi=0$. This is quite unlike the bulk
breathers of the sine-Gordon theory  which approach the vacuum as
their frequencies approach zero. On the other hand,
this behaviour of the boundary breathers is closer to that of
a standard harmonic oscillator of a given frequency whose classical
amplitude may be arbitrarily small. Once quantized, the energy
spectrum of the oscillator depends only on its frequency and the
zero-point energy is there to remind us that an oscillator of
arbitrarily small amplitude is distinct from the vacuum. Similarly, we
expect that the sinh-Gordon boundary breathers will have a
non-zero ground state energy.

For general boundary conditions the properties of the solutions are
less amenable to analysis. However,
numerical investigation of the general boundary breathers indicates that
they are non-singular in the region $x<0$ provided
\begin{equation}\label{aregion}
\cos\frac{\pi
(a_0+a_1)}{2}<0, \qquad
0<\cos\rho <-\cos\frac{\pi (a_0+a_1)}{2}, \qquad
\cos\frac{\pi (a_0-a_1)}{2}>0.
\end{equation}
Again, the breathers have frequencies bounded below because the
parameters are restricted.
We shall take the parameters to
satisfy $a_0\ge a_1$ (since one must be larger than the other
unless they are equal, we take the larger to be $a_0$), and to lie
within the region defined by:
\begin{equation}\label{aranges}
 0\le a_0,a_1\le 1,\quad 0\le a_-\le\frac{1}{2},\quad \frac{1}{2}
\le a_+\le 1,\quad \pi (1-a_+)\le \rho \le \frac{\pi}{2}.
\end{equation}
 As we have noted, the boundary breathers for boundary conditions
preserving the symmetry of the sinh-Gordon equation are included
as the special case $a_0=a_1$.

The energy functional of the sinh-Gordon model incorporating the
 boundary
condition \eqref{bc} is given by
\begin{align}\label{en}
\cal{E}[\phi]=&\int_{-\infty}^0 dx\left(\frac 12\dot{\phi}^2+
\frac 12{\phi'}^2+\frac{2}{\beta^2}
\left(\cosh(\sqrt{2}\beta\phi)-1\right)\right)\nn\\[4pt]
&~~+\frac{2}{\beta^2}\left(\ep_0
(e^{-\frac{\beta}{\sqrt{2}}\phi(0,t)}-1) +\ep_1
(e^{\frac{\beta}{\sqrt{2}}\phi(0,t)}-1)\right),
\end{align}
but it is most easily calculated in terms of the $\tau$ functions
as a boundary term \cite{Del98a},
\begin{equation}\label{enb}
{\cal{E}}[\phi]=\frac{2}{\beta^2}\left.\left(\ep_0
\left(\frac{\tau_+}{\tau_-}-1\right)
+\ep_1\left(\frac{\tau_-}{\tau_+}-1\right)-\left(\frac{\tau_+^\prime}{\tau_+}
+\frac{\tau_-^\prime}{\tau_-}\right)\right)\right|_{\rm x=0}.
\end{equation}
Using this, the energies of the boundary breathers were calculated in
\cite{Cor99} and are
\begin{equation}\label{eb}
{\cal E}=\frac{4}{\beta^2}\left(-2-2\cos\rho + \left(\sin\frac{\pi
a_0}{2} +\sin\frac{\pi a_1}{2}\right)^2\right).
\end{equation}
For the symmetric case with $a_0=a_1$ the energy simplifies to
\begin{equation}
{\cal E}_{\text{breather}}=\frac{8}{\beta^2}(-\cos\rho-\ep).
\end{equation}

In the quantum field theory, the continuum of boundary breather
solutions is expected to lead to a discrete spectrum of boundary
bound states. To estimate this spectrum we shall follow the method
described in \cite{Cor99} which adapts the
techniques developed by Dashen, Hasslacher and
Neveu (DHN) \cite{DHN75}. Although it is not obvious that this method gives an
exact result, it is clear that the result is non-perturbative,
in the sense of being an all orders computation in perturbation theory
in terms of the bulk coupling constant and the boundary parameters.

One of the ingredients to the DHN prescription is the classical action
computed over a single period of the boundary breather. This
quantity is relatively straightforward to calculate for the symmetric
boundary condition but, as far as we can see, is not tractable
analytically for the general breather. Nevertheless, we have
conjectured the result and checked it numerically in two different
ways. The principal difficulty lies in calculating the integral
representing the kinetic energy over a single period.
We maintain this should be given by a simple expression linear in
$a_+$, that is
\begin{equation}\label{KE}
I=\int_0^Tdt \int_{-\infty}^0 dx\, \dot{\phi}^2= \frac{8\pi}{
\beta^2} (\rho -\pi (1-a_+)),
\end{equation}
where $T=\pi/\sin\rho$ is the period of the general breather.
When $a_0=a_1$, eq\eqref{KE} agrees with the result given in
\cite{Cor99};
using Maple it is possible to check that the integral is independent
of $a_-$ by differentiating the integrand with respect to $a_-$
and calculating the resulting integral numerically, obtaining $0$;
it is possible to check directly, again by numerical integration
within the ranges of parameters \eqref{aregion}), that $I$
depends linearly on both $\rho$ and $a_+$. Some details of these computations
will be given in appendix A. The conjectured
expression \eqref{KE} seems to us to be an
astonishing result for which we are quite unable to find an analytical
derivation, despite its beguiling simplicity.

\section{The boundary bootstrap}

\label{s:boot}

For certain ranges  of the parameters $E$ and $F$ the
particle reflection amplitude \eqref{br} has simple poles at
particular imaginary values of $\theta$ on the physical strip,
$0<\im{\theta}<\pi/2$. These are expected to be due to the propagation of
virtual excited boundary states although there are also other
potential explanations via
 generalized Coleman-Thun mechanisms \cite{dorey98, dorey00}. The reason for
this
is the following. Once the boundary condition breaks the bulk symmetry
of the model,
the lowest energy field configuration is no longer $\phi=0$. This
means that in the bulk there will be effective odd-point couplings in
addition to the standard even-point couplings \cite{Chen00} and these
may be used to construct loop  Feynman diagrams `attached' to the
boundary, some of which  may generate poles in the reflection factor
$K_q$.  This possibility
is difficult to analyze and we are unable to pursue it here. We  shall
simply assume the poles are due to virtual excited boundary bound
states without concerning ourselves with their dynamical origin.
The amplitudes
for the reflection of the
sinh-Gordon particle from these excited boundary states are
obtained by the boundary bootstrap \cite{Ghosh94a,
Fring95, Cor94a}. When the reflection factor \eqref{br} has a pole at
$\theta=i\psi$ with $0<\psi<\pi /2$ then the reflection factor
corresponding to the associated excited boundary state is
calculated via the relation
\begin{equation}\label{bbs}
K_1 (\theta )= K_0(\theta ) S(\theta - i\psi )S(\theta + i\psi),
\end{equation}
where $S(\theta)$ is the two-particle S-matrix \eqref{s} and
$K_0(\theta)$
is the ground state reflection factor. Also,
since energy is conserved, the energy of the excited boundary
state relative to the ground state  is given by
\begin{equation}\label{excite} {\cal E}_1={\cal
E}_0 + m(\beta)\cos\psi ,
\end{equation} where $m(\beta )$ is the
mass of the sinh-Gordon particle.

As a parenthetical remark it is worth mentioning the consequences of
this bootstrap procedure for a free, real scalar field of mass $m$
with a linear
boundary condition at $x=0$:
\begin{equation}
 \left. \partial_x\phi\right|_0=-m\lambda\phi .\nn
\end{equation}
Its $S$ matrix is unity but the reflection factor is given by
\begin{equation}
K=\frac{i\sinh\theta +\lambda}{i\sinh\theta -\lambda}=
-\frac{1}{(1+2a)(1-2a)},\nn
\end{equation}
where $\lambda =\cos a\pi$ in the second formula. This has a pole
which indicates (provided $-m<\lambda<0$ or $1>a>1/2$) a
boundary breather of a fixed frequency $\omega$, given  by
$\omega^2=m^2-\lambda^2=m^2\sin^2 a\pi$.
Explicitly, the appropriate normalizeable
solution to the
Klein-Gordon equation and the boundary condition is $\phi=Ae^{-\lambda
x}\cos\omega t$. Repeated application of the bootstrap equations leads
to a tower of boundary states, each with the same reflection factor,
but with energies given by
\begin{equation}
{\cal E}_n={\cal E}_0 + n\omega. \nn
\end{equation}
This is just as one would expect for a harmonic oscillator attached to the
boundary although an alternative dynamical argument would be needed to
determine the ground state energy ${\cal E}_0=\omega/2$.

Returning to the sinh-Gordon case,
let us begin by considering the classical reflection factor
\eqref{bclass}
in the light of the boundary breather parameter restrictions
\eqref{aregion}. It is immediately clear that because of the
restrictions on the parameters only one of the four factors
in the denominator of \eqref{bclass},
namely $(1-a_0-a_1)$ has a pole in the physical strip,
(remember, we have taken the principal values $1>a_+>1/2, \
1/2>a_->0$).
We may also recall the perturbative result reported in \cite{Chen00}
which implies that at least for sufficiently small coupling
$E$ and $F$ do not roam far from their classical values. Bearing in
mind the classical limits \eqref{EFlimits}, this remark
suggests that we may consider the poles associated with $E$ alone,
ignoring $F$.

The remainder of the discussion in this section follows closely the
calculations reported in \cite{Cor99} but, for the sake of
completeness
and small changes of notation,
they will be repeated here. The  regions in $E$  where the amplitude
\eqref{br} has poles on the physical strip are
\begin{equation}\label{Eranges}
{\rm I}: 2>E>1 \quad \hbox{and} \quad {\rm II}: -2<E<-1;
\end{equation}
since  $0\le B\le 2$, the other factors  never have
poles in the physical strip. In region I,  $\psi=\pi (E-1)/2$ and,
using \eqref{bbs}, we derive the reflection factor for the first
excited state,
\begin{equation}\label{firstK}
K_1= \frac{K_S}{(1-E)(1+E)}\,
\frac{(1+E+B)(1-E-B)}{(1-E+B)(1+E-B)},
\end{equation}
where the `spectator' factors have been lumped together in
\begin{equation}
K_S=\frac{(1)(1+B/2)(2-B/2)}{(1-F)(1+F)}.
\end{equation}

The reflection factor \eqref{firstK}
in turn has a new pole at $\psi=\pi (E-1-B)/2$, provided
$B<E-1$, indicating another excited state whose reflection factor
is
\begin{equation}
K_2= \frac{K_S}{(1-E+B)(1+E-B)}\,
\frac{(1+E+B)(1-E-B)}{(1-E+2B)(1+E-2B)}.
\end{equation}
Continuing the procedure leads to a set of excited states with
associated reflection factors given by,
\begin{equation}\label{Kn}
K_n= \frac{K_S}{(1-E+(n-1)B)(1+E-(n-1)B)}\,
\frac{(1+E+B)(1-E-B)}{(1-E+nB) (1+E-nB)}.
\end{equation}
Note, the pole corresponding to the ($n+1$)st state will be
within the correct range provided $E$ satisfies $ 2>E>1+nB $.
Thus, for a given $E$ and $B$ there can be at most a finite number
of bound states, and possibly none. Note too that the reflection
factor for scattering from the $n$th bound state also contains a
pole corresponding to the ($n-1$)st bound state. Since  the factor
$K_S$ is merely a spectator, at no stage  will poles
with $F$-dependent positions enter the game if they did not do so
at the start.

The energies ${\cal E}_n$ of the boundary states are found by repeatedly
applying \eqref{excite}. They  are  given by,
\begin{equation}\label{ediff}
{\cal E}_{n+1}={\cal E}_n + m(\beta )\cos \frac{\pi}{2} (nB-E+1).
\end{equation}
This is the result that we will compare with the quantization
of the classical breather spectrum in order to determine how
$E, F$ and $m(\beta)$ depend upon $a_0$ and $a_1$.

The poles in region II do not represent a new set of states. In
the discussion section we shall make some comments concerning the
relative r\^oles of $E$ and $F$.

\section{Semi-classical quantization}

\label{s: semi}

The  first step in carrying out the semi-classical calculation is
to solve the sinh-Gordon equation linearised in the presence of
the boundary breathers. Setting $\phi =\phi_0+\eta$, where $\phi_o$ is
a classical breather, the linear
wave equations which ought to be satisfied by the fluctuations are:
\begin{equation}\label{linear}
\frac{\partial^2\eta}{\partial t^2}-\frac{\partial^2\eta}{\partial
x^2} +4 \eta \cosh\sqrt{2}\beta\phi_0 =0, \qquad
\left(\frac{\partial \eta}{
\partial x}+  \eta\left[\ep_0e^{-\beta\phi_0/\sqrt{2}}+
\ep_1e^{\beta\phi_0/\sqrt{2}}\right]\right)_{\rm
x=0}=0.
\end{equation}
It is convenient to solve \eqref{linear} by perturbing \eqref{st}.
In other words, we may take
\begin{equation}
\eta=\frac{\tau_-\delta\tau_+-\tau_+\delta\tau_-}{ \tau_+\tau_-},
\end{equation}
with $\delta\tau_\pm$ chosen in turn by adding a pair of `small'
exponential terms to \eqref{taupm} as follows. Take $e_1$ and $e_2$
defined by
\begin{equation}
e_1=\lambda_1\, e^{-i\omega t +ikx}, \ \ e_2=\lambda_2\,
e^{-i\omega t -ikx},\ \ \ \ \ \omega^2-k^2=4, \nn
\end{equation}
where $\lambda_1, \lambda_2$ are infinitesimally small.  We use
Hirota's method again, but with five basic ingredients this time,
instead of the previous three, keeping only the terms linear in $e_1$
and $e_2$.
Thus, we write,
\begin{eqnarray}\label{deform}
\delta\tau_\pm&=&\sum_{p=1,2} e_p \bigg(
\pm 1 +(\mu_{p1}E_1+\mu_{p2}E_2+\mu_{p3}E_3)\nonumber\\ [0pt] &
& \ \ \ \ \ \ \ \ \ \ \ \ \   \pm
 (\mu_{p1}\mu_{p2}A_{12}E_1E_2+\mu_{p1}\mu_{p3}A_{13}E_1E_3+
\mu_{p2}\mu_{p3}A_{23}E_2E_3)\nn \\[8pt]
&&\ \ \ \ \ \ \ \ \ \ \ \ \ \ \ \ \ \ \ \ \ \ \ \ \ \
  + \mu_{p1}\mu_{p2}\mu_{p3}A_{12}A_{13}A_{23}
E_1E_2E_3\bigg).
\end{eqnarray}
The new coefficients $\mu_{pq}$ are obtained from the general formulae
\eqref{taupieces} and given by:
\begin{eqnarray}
 \mu_{11}&=&\frac{1}{\mu_{22}}=\frac{ik\cos\rho -\omega\sin\rho -2}
{ik\cos\rho -\omega\sin\rho +2},\ \ \mu_{12}=\frac{1}{\mu_{21}}=
\frac{ik\cos\rho+\omega\sin\rho -2 }{ik\cos\rho+\omega\sin\rho
+2},\\ [4pt] \nn
&& \ \ \ \ \ \ \ \ \ \ \ \ \ \ \ \ \ \ \ \ \ \ \
\mu_{13}=\frac{1}{\mu_{23}}=\frac{ik-2}{ik+2}.
\end{eqnarray}
 Matching
the boundary condition at $x=0$ fixes the ratio
$\lambda_2/\lambda_1$ to be \footnote{\ Remark: the right hand side of
the corresponding result in \cite{Cor99}, eq(5.5), has a misprint
and the printed formula ought to be inverted to obtain the correct
result for $K_B$}
\begin{equation}\label{lratio}
K_B=\frac{\lambda_2}{\lambda_1}=\left(\frac{ik-2\cos\rho}{
ik+2\cos\rho}\right)^2\ \frac{ik-2}{ ik+2}\ \frac{ik-2c_+}{ik+2c_+}\
 \frac{ik+2c_-}{ik-2c_-},
\end{equation}
where $c_\pm =\cos \pi(a_0\pm a_1)/2$. The result \eqref{lratio} is
not intended to be obvious; its derivation relied heavily on using a
symbolic algebraic manipulation language---in our case, Maple.
In the limit $x\rightarrow -\infty$, the fluctuation $\eta$ is a
superposition of left and right moving plane waves,
\begin{equation}\label{breflect}
\eta\sim \lambda_1\, e^{-i\omega t}\left(e^{ikx}+K_B\,
e^{-ikx}\right) ,
\end{equation}
and the relative phase of these waves
defines the classical reflection factor corresponding to the
boundary breather. Taking $\cos\rho = -c_+$, the breather
collapses to the static  ground state solution  and the reflection
factor collapses to
\begin{equation}\label{vreflect}
K_0=\frac{ik-2}{ ik+2}\ \frac{ik+2c_+}{ik-2c_+}\
 \frac{ik+2c_-}{ik-2c_-},
\end{equation}
which, with $k=2\sinh\theta$, is simply an alternative form (ie without
using the `block' notation) of the expression
 \eqref{bclass}.

The period $T=\pi /\sin\rho$ of the boundary breather defines the
`stability angles' via
\begin{equation}\label{stab}
\eta (t+T,x)=e^{-i\nu }\eta(t,x)\equiv e^{-i\omega T}\eta(t,x)
\end{equation}
and the field theoretical version of the WKB approximation makes
use of the stability angles together with a regulator to calculate
a quantum action. If there are no boundaries, it is natural to add
some artificially to render the spectrum discrete and facilitate
the necessary calculations. One way, the simplest and most
commonly
used, would be to place the
field theory in an interval $[-L,L]$ with periodic boundary
conditions and to manipulate the sum over the discrete stability
angles. However, since we have one boundary already prescribed, we
need
to do something else as suggested in \cite{Cor99}.
It is convenient to treat the sinh-Gordon model in
the interval $[-L,0]$ and to impose the Dirichlet condition
$\eta(t,-L)=0$. Since the limit $L\rightarrow\infty$ will  be
taken eventually, the stability angles for the boundary breather
($\nu_B$), or the vacuum solution ($\nu_0$) are effectively
determined by the reflection factors given in \eqref{breflect} or
\eqref{vreflect}, respectively. A potentially more interesting
calculation
but one which we do not yet have the machinery to carry out, would be
to consider two sets of two-parameter boundary conditions. However,
the breathers we are considering are inadequate for that.

Following \cite{DHN75, Raj82} we need to calculate a sum over the
stability angles and use it to correct the classical action. Thus,
\begin{equation}\label{D}
\Delta=\frac{1}{ 2}\sum \, (\nu_B-\nu_0)\equiv \frac{T}{ 2}\sum\,
\left(\sqrt{ k_B^2+4}-\sqrt{k_0^2+4}\right), \end{equation}
where  $k_B$ and $k_0$ are the sets of (discrete) solutions to
\begin{eqnarray}\label{qconditions}
&& e^{2ik_BL}=-\,
\left(\frac{ik_B+2\cos\rho}{ ik_B-2\cos\rho}\right)^2\
\frac{ik_B+2}{ ik_B-2}\ \frac{ik_B+2c_+}{ik_B-2c_+}\
\frac{ik_B-2c_-}{ik_B+2c_-},\nn\\  [4pt]
&&\hspace{1cm} e^{2ik_0L}=-\,
\frac{ik_0+2}{ ik_0-2}\ \frac{ik_0-2c_+}{ik_0+2c_+}\
\frac{ik_0-2c_-}{ik_0+2c_-} .
\end{eqnarray}
Once $\Delta$ is known, the quantum action is defined by
\begin{equation}\label{quaction}
S_{\rm qu}=S_{\rm cl}-\Delta,
\end{equation}
where the classical action is readily calculated from the kinetic
energy integral \eqref{KE} and the total energy \eqref{eb}:
\begin{equation}\label{claction}
S_{\rm cl}=\int_0^Tdt\int_{-\infty}^0dx\, {\cal L}=
\frac{8\pi}{\beta^2}\left(\rho- \pi (1-a_+)  +\frac{1}{\sin\rho}
\left(\cos\rho +1 -\frac{1}{2}(1-c_+)(1+c_-)\right)\right).
\end{equation}

We shall proceed along the lines described in \cite{Cor99}
noting that for large $k$ the solutions to
either of \eqref{qconditions} are close to $$k_n=\left(n+\frac{1}{
2}\right)\frac{\pi}{ L},$$ and so it is reasonable to set
$(k_B)_n=(k_0)_n+\kappa((k_0)_n) /L$ where, for $L$ large, the
function $\kappa$ is given approximately by
\begin{equation}\label{kappadef}
e^{2i\kappa (k)}=\left(\frac{ik+2\cos\rho}{ ik-2\cos\rho}\
\frac{ik+2c_+}{ ik-2c_+}\right)^2\, .
\end{equation}
Interestingly, all dependence upon $c_-$ has dropped out
and therefore, from this point on, the calculation is
identical to the corresponding part of the calculation presented in
\cite{Cor99} with $\ep=\cos\pi a$ replaced by
$c_+$.

In terms of $\kappa$ the expression \eqref{D} is rewritten
\begin{equation}
\Delta\sim \frac{T}{ 2L}\sum_{n\ge 0}\, \frac{(k_0)_n\kappa
((k_0)_n)}{ \sqrt{ (k_0)_n^2+4}}+O(1/L^2),\nonumber \end{equation}
and this, in turn, as $L\rightarrow\infty$ can be converted to a
convenient (but actually divergent) integral,
\begin{equation}\label{Da} \Delta =\frac{T}{
2\pi}\int_0^\infty\, dk \frac{k \kappa(k)}{ \sqrt{k^2+4}}
\end{equation}
with which we shall have to deal.  Note that $\kappa$ vanishes
when $\cos\rho = -c_+$.
Integrating \eqref{Da} by parts we find
\begin{equation}\label{Db}
\Delta=\frac{T}{ 2\pi}\left( \left. \kappa \sqrt{k^2+4}\,
\right|^\infty_0 -\int_0^\infty dk \frac{d\kappa}{ d k}
\sqrt{k^2+4}\right),
\end{equation}
where
\begin{equation}\label{kderv}
\frac{d\kappa}{dk} = \frac{4\cos\rho}{ k^2+4\cos^2\rho} +
\frac{4c_+}{ k^2+4c_+^2},
\end{equation}
and we note that with a suitable choice of branch
\begin{equation}\label{klimit}
\kappa\sim -\frac{4\cos\rho}{ k}-\frac{4c_+}{ k}\ \
\hbox{as}\ \ k\rightarrow\infty .
\end{equation}
Using \eqref{klimit} and recalling that $\cos\rho < -c_+$, we
deduce that $\kappa$ approaches zero from above as $k\rightarrow
\infty$. Also, from \eqref{kderv} it is clear that the derivative
of $\kappa$ is positive  near $k=0$ but negative as $k\rightarrow
\infty$. Hence, the first term in \eqref{Db} is well-defined and
the appropriate branch of $\kappa$ has $\kappa (0)=0$. On the
other hand, the derivative of $\kappa$ is not decaying
sufficiently rapidly to ensure the second term in \eqref{Db} is
finite. However, this was to be expected since a perturbative
analysis of the sinh-Gordon model confined to a half-line needs
mass and boundary counter terms to remove logarithmic divergences
(which would be removed automatically by normal-ordering the
products of fields in the bulk theory). With this in mind, the
integral remaining in \eqref{Db} should be replaced by
\begin{equation}\label{counter}
\int_0^\infty dk\sqrt{k^2+4}\left(\frac{4\cos\rho}{
k^2+4\cos^2\rho}- \frac{4\cos\rho}{ k^2+4}+\frac{4c_+}{
k^2+4c_+^2}- \frac{4c_+}{ k^2+4}\right),
\end{equation}
the first counter-term removing the bulk divergence and the second
being there to remove a similar divergence associated with the
boundary. In effect, we are regarding the parameter $a$ as
describing the bare coupling which appears in the boundary part of
the Lagrangian once it is written in terms of normal-ordered
products of fields. The counter-terms clearly respect the symmetry
and the whole expression vanishes when $\rho = \pi(1- a_+)$. The
integrals in \eqref{counter} need to be treated carefully noting
that $\cos\rho >0$ but $c_+ <0$.

Besides the towers of real solutions to \eqref{qconditions},
there is also a discrete set of solutions for which $k_0$ and $k_B$
are pure imaginary. These were not discussed in
\cite{Cor99} but including them in the argument would not have
altered the
conclusions, as we shall show. First, notice that as $L\rightarrow
\infty$ the imaginary solutions which survive are either  zeroes or
poles of the right hand sides of the equations in \eqref{qconditions},
according to the signs of $ik_0$ or $ik_B$. Thus for $k_0$ we have
$ik_0=2,-2c_+$if $ik_0>0$ and $ik_0=-2, 2c_+$ if $ik_0<0$, while for $k_B$
we have $ik_B=2\cos\rho, 2$ if $ik_B>0$ and $ik_B=-2\cos\rho, -2$ if
$ik_B<0$. However, the signs should be disregarded because for either
sign each solution for a specific value of $|ik_0|$ or $|ik_B|$
represents a
single `bound-state' function $\eta$. Taking this into account, these
special solutions contribute to
$\Delta$ an additional piece:
\begin{equation}\label{extra}
 T(\sin\rho - \sin\pi a_+)=\pi - \frac{\pi \sin\pi a_+}{\sin\rho}.
\end{equation}

Assembling the various components leads to
\begin{equation}\label{Dc}
\Delta=\pi  -\frac{2}{ \sin\rho}\left(\cos\rho +\cos\pi a_+ +\rho
\sin\rho +\pi(a_+ -1/2)\sin\pi a_+ \right).
\end{equation}
 Recalling \eqref{claction}, and using \eqref{Dc}, the
quantum action defined in \eqref{quaction} is given by an
expression of the form
\begin{equation}\label{qactiona}
S_{\rm qu}=\frac{4}{ B}\left(\frac{\cos\rho}{\sin\rho} +\rho
-\frac{\pi}{ 2} \right) +\frac{8\pi}{\beta^2}\left(\pi a_+
-\frac{\pi}{ 2}\right) +\frac{\Gamma(a_+,a_-)}{ \sin\rho} ,
\end{equation} where $\Gamma$ is
independent of $\rho$,
\begin{equation}\label{Gamma}
\Gamma(a_+,a_-)=2\pi\left(-\frac{2}{\beta^2}(1-c_+)
(1+c_-) +\frac{4}{\beta^2}+\frac{c_+}{\pi} +
(a_+-1/2)\sin \pi a_+\right). \nn
\end{equation}

Once the quantum action is determined, the quantum energy is
defined by
\begin{eqnarray}\label{qenergy}
{\cal E}_{\rm qu}&=& -\frac{\partial S_{\rm qu}}{ \partial
T}=\frac{\sin^2\rho}{ \pi \cos\rho}\frac{\partial S_{\rm
qu}}{\partial\rho} =-\frac{4}{ \pi B}\cos\rho -\frac{\Gamma  (a_+,a_-)}
{\pi},
\end{eqnarray}
and the WKB quantization condition states that
\begin{equation}\label{WKB} W_{\rm qu}=
S_{\rm qu}+T{\cal E}_{\rm qu}= \frac{4}{ B} \left(\rho -\frac{\pi
}{ 2}\right) +\frac{8\pi}{\beta^2}\left(\pi a_+- \frac{\pi}{
2}\right) =2n\pi .
\end{equation}
Where, $n$ is  a positive integer or zero.\footnote{ Including the
imaginary solutions of \eqref{qconditions} obviates the need for making
the change $n\rightarrow n+1/2$, since we shall see below the
zero-point energy is automatically correct. In this sense, including
the
imaginary solutions leads more naturally to the expected result.}
Hence, the energies of the quantized
boundary breather states are
 determined by a set of special angles $\rho_n$,
\begin{equation}
\rho_n=\frac{\pi}{2}\left(1+nB -
\frac{2\pi B}{\beta^2} (2a_+ -1)\right),
\end{equation}
and given by
\begin{equation}\label{qspectrum} {\cal E}_n=-\frac{4}{ \pi B}\cos
\rho_n - \frac{\Gamma}{ \pi} = -\frac{4}{ \pi B} \cos \frac{\pi}{
2}\left(nB + 1 -\frac{2\pi B}{ \beta^2}
(2a_+-1)\right) -\frac{\Gamma}{ \pi}.
\end{equation}
Notice that as $\beta\rightarrow 0$, $\rho_n\rightarrow \pi (1-a_+)$
independently of $n$. Thus, the frequencies collapse to the lowest
allowed frequency, namely $\omega_0=2\sin \pi a_+$. On the other
hand, in the same limit the energies are independent of $\beta$
and non-zero,
\begin{equation}\label{limitspectrum}
{\cal E}_n\rightarrow \left(n+\frac{1}{2}\right)\omega_0.
\end{equation}
This is precisely the spectrum of a harmonic oscillator vibrating
at the fundamental frequency $\omega_0$.

Using \eqref{qspectrum} the corresponding differences in the
energy levels are given by
\begin{equation}\label{qediff}
{\cal E}_{n+1}={\cal E}_{n} +\frac{8}{ \pi B}\sin \frac{\pi B}{
4}\, \cos \frac{\pi}{ 2}\left(\frac{2\pi
B}{\beta^2}(2a_+-1)-\left(n+\frac{1}{2}\right)B\right).
\end{equation}
Comparing \eqref{qediff} with the outcome of the bootstrap
calculation \eqref{ediff} allows us to identify the
 parameter $E$ which appeared in the expression for the
reflection factor \eqref{br}. Thus, from the first excited
level we deduce,
\begin{equation}\label{E}
E(a_0,a_1,\beta)=  \frac{2\pi B}{\beta^2}(2a_+-1)-\frac{1}{2} B+1
\equiv 2a_+
\left(1-\frac{B}{ 2}\right)= (a_0+a_1)\left(1-\frac{B}{ 2}\right),
\end{equation}
and then all the other levels match up without further restriction.
This is in agreement with the suggestions made in \cite{Chen00}.

As was pointed out previously in \cite{Cor99},
the  comparison with \eqref{ediff} also permits us to deduce an
expression for $m(\beta)$, the mass of the sinh-Gordon particle:
\begin{equation}\label{mass}
m({\beta})=\frac{8}{ \pi B}\sin \frac{\pi B}{ 4}.
\end{equation}

\section{Discussion}

\label{s:discuss}

\phantom{\hspace{0pt}}From the point of view of the classical
reflection factor
\eqref{bclass},
it is clear that the parametrisation \eqref{ea} is the
most appropriate. Moreover, the expressions for $\ep_0$ and $\ep_1$ in
terms of $a_0$ and $a_1$ are invariant under independent changes of
signs of either $a_0$ or $a_1$ and this symmetry is incorporated in the
factors appearing in the denominator of \eqref{bclass}. We expect
the symmetry under reversing the signs of $a_0$ and $a_1$ should
persist in the quantum reflection factor, and indeed that was part of
the thinking behind the conjectured forms for $E$ and $F$ presented in
\eqref{EFconjecture}. In other words, having calculated $E$ in
\eqref{E}  we can immediately deduce the partner expression for $F$.

We found that the general breathers exist as non-singular real
solutions provided the parameters satisfy the constraints
\eqref{aranges}. It is gratifying to discover at the end of the
calculation
that the renormalization factor $1-B/2$ in \eqref{EFconjecture}
for real  coupling $\beta$ is restricted to lie in the range
\begin{equation}
0\le 1-\frac{B}{2}\le 1, \nn
\end{equation}
and thus has the effect of scaling the parameters but preserving the
constraints \eqref{aranges}. Notice that with $E$ and $F$ given by
\eqref{EFconjecture} it is impossible to have $E$ and $F$ lying
simultaneously in regions \eqref{Eranges} for which the bound state
poles of the reflection factor lie on the physical strip. Thus,
the poles depending on $E$ or $F$ are alternatives and actually
lead to the same tower of states simply seen differently in terms of
the parameters $a_0$ and $a_1$.

Changing the sign of either $a_0$ or $a_1$ takes us outside the
principal ranges of these parameters and obliges us to reformulate the
breather solutions. In fact, there are no others besides the ones we
have found although the expressions for them will be a little
different if we adopt a different principal region.

It is worth emphasizing that the expressions \eqref{EFconjecture}
incorporate a weak-strong coupling duality \cite{Chen00}
which extends that enjoyed by the S-matrix itself. If a new
triple of coupling constants $(a_0^*,a_1^*,\beta^*)$ is defined by
\begin{equation}
(a_0^*,a_1^*,\beta^*)=\frac{4\pi}{\beta^2}(a_0,a_1,\beta),\nn
\end{equation}
then
\begin{equation}
E(a_0^*,a_1^*,\beta^*)=\frac{4\pi}{\beta^2}(a_0+a_1)\frac{B}{2}\equiv
(a_0+a_1)(1-B/2)=E(a_0,a_1,\beta),
\end{equation}
and similarly for $F$.

We must acknowledge that the semi-classical
approximation, although non-perturbative, is not guaranteed to be
exact. However, for the bulk sine-Gordon model, the Dashen, Hasslacher and
Neveu approach does give exact information concerning the spectrum
and we would expect that the same should be true here. Unfortunately,
alternative exact computations of the spectrum are not yet available
for comparisons to be made.

As far as other models are concerned, it will be interesting to try
out these methods in other cases where the boundary parameters form a
continuous set. The simplest such example, in which there are two
distinct one-parameter families of boundary parameters, is the model
based on the $a_2^{(2)}$ root data \cite{Bow95}.

\vskip 1cm

\noindent{\bf Acknowledgements}
\vskip .5cm \noindent One of us (AT) is
supported by a Leverhulme Fellowship, the other (EC)
 thanks the University of
Durham for a Visiting Professorship and the Yukawa Institute for
 Theoretical Physics for hospitality during a visit funded by a Royal
 Society/JSPS exchange programme;  both of us have been
partially supported by a TMR Network grant of the European
Commission contract number FMRX-CT-960012 and we thank the
 Universities of Montpellier and Mons for their hospitality.
 We are indebted to Nigel Glover
for advice on using  Maple and to Gustav Delius for discussion and
comments.

\appendix
\section{Appendix}

In this appendix we shall give some information concerning the kinetic
energy integral \eqref{KE}. The first step in evaluating the integral
is to perform the integration over time. This
is relatively easy and leads to:
\begin{equation}\label{integrand}
I(x)=\int_0^Tdt\dot{\phi}^2=\frac{8\pi\sin\rho}{\beta^2}
\left[\frac{2C\sqrt{B^2-A^2}}{AD-BC} +\frac{B}{\sqrt{B^2-A^2}}-
\frac{2A\sqrt{D^2-C^2}}{AD-BC} +\frac{D}{\sqrt{D^2-C^2}}\right],
\end{equation}
where the right hand side is built from the components $A,B,C,D$
which are defined by
\begin{eqnarray}
&&\hspace{0cm}
A=\frac{su}{q}(1-q^2)(1-rv),\ C=-\frac{su}{q}(1-q^2)(1+rv),\nn \\
&&B=\frac{1}{q^2}\left((rv-s^2q^2u^2)+q^2(1-s^2q^2r vu^2)\right),\nn \\
&&D=\frac{1}{q^2}\left(-(rv+s^2q^2u^2)+q^2(1+s^2q^2r vu^2)\right),
\end{eqnarray}
with
\begin{equation}
u=e^{2x\cos\rho},\ v= e^{2x}.
\end{equation}
The parameters $q,r,s$ were defined previously in \eqref{cdaliter}.
Notice, when $a_0=a_1$ we have $r=0$ allowing all of  these expressions
to simplify dramatically. Consequently, the integral of the right hand
side of
\eqref{integrand}
can be done by making a suitable change of variables. Thus, for $a_0=a_1=a$
we find:
\begin{equation}
\int_{-\infty}^0 dx\, I(x)=\frac{8\pi}{\beta^2}(\rho -\pi(1-a)).
\end{equation}
 However, in the general
case we have not found a change of variables which simplifies
\eqref{integrand}. Therefore, we have had to proceed numerically.

One check we have made is to differentiate \eqref{integrand} with
respect to $a_-$ and demonstrate
\begin{equation}
\frac{\partial}{\partial a_-}\int_{-\infty}^0 dx\, I(x) =0,
\end{equation}
provided $a_\pm$ lie within the ranges \eqref{aranges}. Another is
to calculate the integral directly as a function of $a_+$ and $\rho$
and demonstrate it depends linearly on each of these parameters
separately.
The following plots generated by Maple indicate unambiguously the
linear dependence on $a_+$ (Figure 1)
  \begin{center}
 \makebox{ \includegraphics{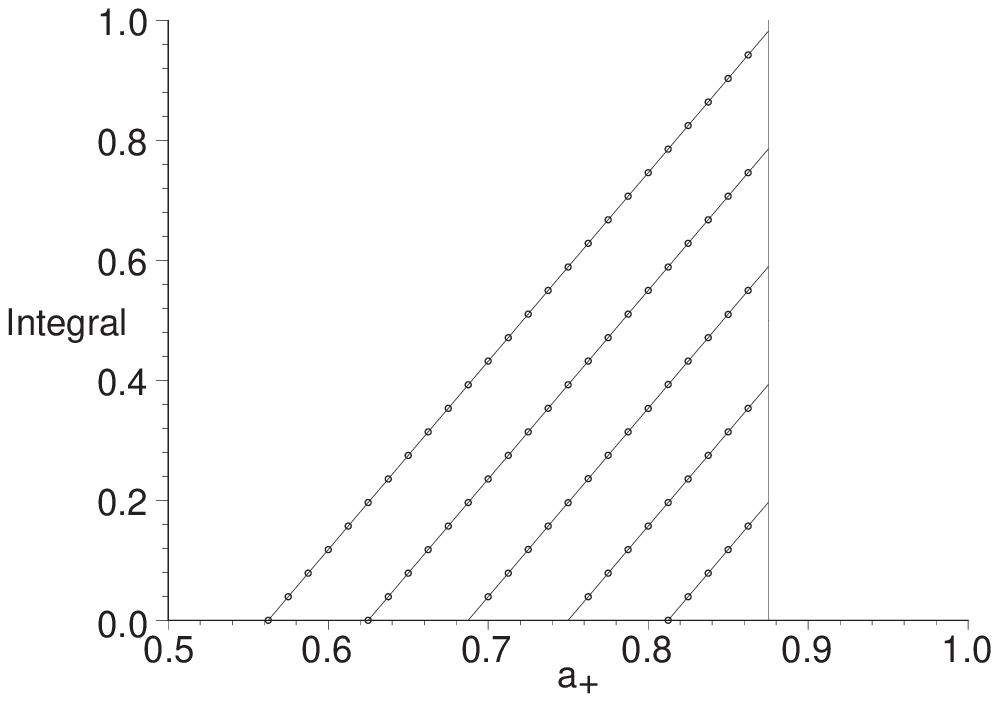}}
  \end{center}

  \begin{center}
  Figure 1: The integral as a function of $a_+$.
  \end{center}
or on $\rho$ (Figure 2)
 \begin{center}
 \makebox{ \includegraphics{fig2.eps}}
  \end{center}

  \begin{center}
  Figure 2: The integral as a function of $\rho$.
  \end{center}
In Fig(1) we have overlaid several plots. The dots represent the
numerical values of the integral as $a_+$ varies from $1-\rho/\pi$ to $1-a_-$,
in accordance with \eqref{aranges}, for each of five different
values of $\rho$ (viz. $\rho/\pi = (2+k)/16, \ k=1,\dots , 5)$;
the continuous line is simply the conjectured value of
the integral given in \eqref{KE} plotted as a function of $a_+$ for the same
five values of $\rho$. In all plots, the factor $8\pi/\beta^2$ has been ignored.
The vertical line indicates the upper bound on $a_+$ for $a_-=1/8$.
 Fig(2) presents similar plots for the integral as
a function of $\rho/\pi$ in the range $1-a_+$ to $1/2$ for five different
values of $a_+$
(viz. $a_+=(8+k)/16, \ k=1,\dots , 5)$.
We regard the
sets of parallel lines depicted in Figs(1,2)
as very convincing numerical evidence for \eqref{KE}.


\begin{thebibliography}{99}

\bibitem{Bow95} P. Bowcock, E. Corrigan, P. E. Dorey and R. H. Rietdijk,
\textit{Classically integrable boundary conditions for affine Toda
 field  theories}, Nucl. Phys. {\bf B445} (1995) 469;   \hepth{9501098}.

\bibitem{Bra90} H.W. Braden, E. Corrigan, P.E. Dorey and R. Sasaki,
\textit{Affine Toda field theory and exact S-matrices},
\npb{338}{1990} {689}.

\bibitem{Chen00}  A. Chenaghlou and E. Corrigan, \textit{First order quantum
corrections to the classical reflection factor of the sinh-Gordon model}, to
appear in Int. J. Mod. Phys. {\bf A}; \hepth{0002065}.

\bibitem{Cor94a} E. Corrigan, P.E. Dorey, R.H. Rietdijk and R. Sasaki,
\textit{Affine Toda field theory on a half line},
\plb{333}{1994}{83}; \hepth{9404108}.

\bibitem{Cor95} E. Corrigan, P.E. Dorey and R.H. Rietdijk,
\textit{Aspects of affine Toda field theory on a half-line}, Prog.
Theor. Phys. Suppl. {\bf 118} (1995) 143; \hepth{9407148}.

\bibitem{Cor99}  E. Corrigan and G Delius, \textit{Boundary breathers in
the sinh-Gordon model}, J. Phys. {\bf A32} (1999) 8001-14; \hepth{9909145}.

\bibitem{Del98a} G.W. Delius, \textit{Restricting affine Toda theory
to the half-line}, J. High Energy Phys. {\bf 09} (1998) 016;
\hepth{9807189}.

\bibitem{DHN75} R.F. Dashen, B. Hasslacher and A. Neveu,
\textit{The particle spectrum in model field theories from
semi-classical functional integral techniques}, Phys. Rev. {\bf
D11} (1975) 3424.

\bibitem{dorey98} P.E.\ Dorey, R.\ Tateo and G. Watts, {\em
Generalisations of the Coleman--Thun mechanism and boundary
reflection factors},  \plb{448}{1999}{249}; \hepth{9810098}.

\bibitem{Fadd78}
L.D. Faddeev and V.E. Korepin, {\it Quantum theory of solitons},
Phys. Rep. {\bf 42} (1978) 1-87.

\bibitem{Fring95} A. Fring and R. K\"oberle, \textit{Boundary
bound states in affine Toda field theories}, Int. J. Mod. Phys.
{\bf A10} (1995) 739; \hepth{9404188}.

\bibitem{Ghosh94a} S.\ Ghoshal and A.\ Zamolodchikov, {\em
Boundary $S$-Matrix and Boundary State in Two Dimensional
Integrable Field Theory}, Int. Jour. Mod. Phys.\ {\bf A9} (1994),
3841; \hepth{9306002}.

\bibitem{Ghosh94b} S.\ Ghoshal, {\em Bound State Boundary $S$-Matrix
of the Sine-Gordon Model}, Int. J. Mod. Phys.\ {\bf A9} (1994),
4801; \hepth{9310188}.

\bibitem{Hiro80} R. Hirota, \textit{Direct methods in soliton theory},
in {\it `Solitons'}, eds R.K. Bullough and P.J. Caudrey (Berlin:
Springer 1980).

\bibitem{MacI95} A. MacIntyre, \textit{Integrable boundary conditions
for classical sine-Gordon theory}, J. Phys. {\bf A28} (1995) 1089;
\hepth{9410026}.

\bibitem{dorey00} P. Mattsson and P.E.\ Dorey, {\em Boundary
spectrum in the sine-Gordon model with Dirichlet boundary
conditions}; \hepth{0008071}

\bibitem{Raj82} R. Rajaraman, \textit{Solitons and Instantons}
(North Holland 1982).

\bibitem{Skly89} E.K. Sklyanin, \textit{Exact quantization of the
sinh-Gordon model}, Nucl. Phys. {\bf B326} (1989) 719.

\bibitem{Zam79} A.B. Zamolodchikov and Al.B. Zamolodchikov,
\textit{Factorized S-Matrices in Two Dimensions as the Exact
Solutions of Certain Relativistic Quantum Field Theory
Models},Ann. Phys. 120 (1979) 253-291.

\end{thebibliography}
\end{document}